\documentclass[
 aps,prx, 
 amsmath,amssymb,
 reprint,  
]{revtex4-1}

\usepackage[utf8]{inputenc}
\usepackage{xcolor}
\usepackage{graphicx}
\usepackage{float}
\usepackage{amsmath}
\usepackage{enumitem}
\usepackage{bm}
\usepackage{bbold}
\usepackage{cases}
\usepackage{upgreek}
\usepackage[colorlinks=true, citecolor=magenta, linkcolor=cyan ]{hyperref}
\usepackage{braket}
\usepackage[normalem]{ulem}
\usepackage{lipsum}
\usepackage{dcolumn}  

\newcommand \rmm[1]  { \textrm{#1} }

\makeatletter
\def\@email#1#2{%
 \endgroup
 \patchcmd{\titleblock@produce}
  {\frontmatter@RRAPformat}
  {\frontmatter@RRAPformat{\produce@RRAP{*#1\href{mailto:#2}{#2}}}\frontmatter@RRAPformat}
  {}{}
}%
\makeatother

\begin{document}

\preprint{AIP/123-QED}

\title{Anomalous transport of small polarons arises from transient lattice relaxation or immovable boundaries}

\author{Srijan Bhattacharyya}
\affiliation{Department of Chemistry, University of Colorado Boulder, Boulder, CO 80309, USA}

\author{Thomas Sayer}
\affiliation{Department of Chemistry, University of Colorado Boulder, Boulder, CO 80309, USA} 

\author{Andr\'{e}s Montoya-Castillo}
\homepage{Andres.MontoyaCastillo@colorado.edu}
\affiliation{Department of Chemistry, University of Colorado Boulder, Boulder, CO 80309, USA} 

\date{\today}

\begin{abstract}
Elucidating transport mechanisms and predicting transport coefficients is crucial for advancing material innovation, design, and application. Yet, state-of-the-art calculations are restricted to exact simulations of small lattices with severe finite-size effects or approximate simulations that assume the nature of transport. We leverage recent algorithmic advances to perform exact simulations of the celebrated Holstein model that systematically quantify and eliminate finite-size effects to gain insights into small polaron formation and the nature and timescales of its transport. We perform the first systematic comparison of the performance of two distinct approaches to predict charge carrier mobility: equilibrium-based Green-Kubo relations and nonequilibrium relaxation methods. Our investigation uncovers when and why disparities arise between these ubiquitously used techniques, revealing that the equilibrium-based method is highly sensitive to system topology whereas the nonequilibrium approach requires bigger system sizes to reveal its diffusive region. Contrary to assumptions made in standard perturbative calculations, our results demonstrate that small polarons exhibit anomalous transport and that it manifests transiently, due to nonequilibrium lattice relaxation, or permanently, as a signature of immovable boundaries. These findings have consequences for applications including the utilization of organic polymers in organic electronics and transition metal oxides in photocatalysis.
\end{abstract}

\maketitle

Materials design and their subsequent application have driven scientific and technological innovation for decades. Understanding how a material responds to external factors, such as temperature gradients or mechanical stress, is crucial for effective design principles~\cite{yu2018mechanical, park2016material}. Transport coefficients encode how materials respond to these external stimuli and offer a figure of merit in the search for novel materials: diffusion coefficients of biomolecules for drug delivery~\cite{mitrea2022modulating}, Seebeck coefficients for thermoelectric generators~\cite{mahapatra2017seebeck}, proton conductivity in membrane channels for flow batteries~\cite{xin2021zr}, and electron mobility in organic electronics and biocompatible device design~\cite{hermosa2013intrinsic, bostick2018protein}. Great strides in spectroscopy have recently enabled transport measurements of photogenerated energy carriers with unprecedented detail~\cite{ginsberg2020spatially} and advances in electronic structure theory have transformed our ability to parameterize physically transparent models~\cite{giustino2017electron}. Yet, while progress on the quantitative dependence of transport coefficients on the underlying chemical structure is promising~\cite{stuart2010emerging, kadic20193d}, the challenge is complicated by the difficulty of accurately predicting the quantum dynamics of carriers, which can exhibit either typical or anomalous transport. Indeed, questions of when, how, and over what timescales one should expect anomalous transport to emerge engender intense interest~\cite{klages2008anomalous}. Hence, it is critical to interrogate transport in fundamental models and elucidate their mechanistic dependence on a system's microscopic details~\cite{oberhofer2017charge}. 

Anomalous transport is an important feature across many important processes ranging from biomolecular flow in cells~\cite{hofling2013anomalous}, to nanoparticle coordination~\cite{srivastava2013structure}, gas diffusion through porous materials~\cite{kim2020anomalous}, and even penetrant transport in glassy polymers~\cite{peppas1983anomalous}. The ubiquity of this phenomenon has ignited an intense theoretical focus on uncovering the sources and timescales of anomalous transport~\cite{szymanski2009elucidating, thiel2013disentangling, gopalakrishnan2019kinetic}. Theory and experiment indicate that it requires at least one of the following conditions: heterogeneous environments~\cite{liu2022heterogeneous}, mixed motion with both diffusion and trapping or binding~\cite{berkowitz1997anomalous}, or the presence of memory effects~\cite{weber2010bacterial}. In contrast to standard perturbative approaches that impose typical transport at the level of their rate expressions~\cite{marcus1993electron, holstein1959studies}, here we will leverage numerically exact simulations---of unprecedented reach in system size and time---to establish that polaron transport can be transiently or permanently anomalous and investigate the source of these anomalous behaviors to determine which of the aforementioned conditions are at play in polaron-mediated solid-state charge transport.

\begin{figure*}[!th]
\begin{center} 
\vspace{-20pt}
    \resizebox{.45\textwidth}{!}{\includegraphics{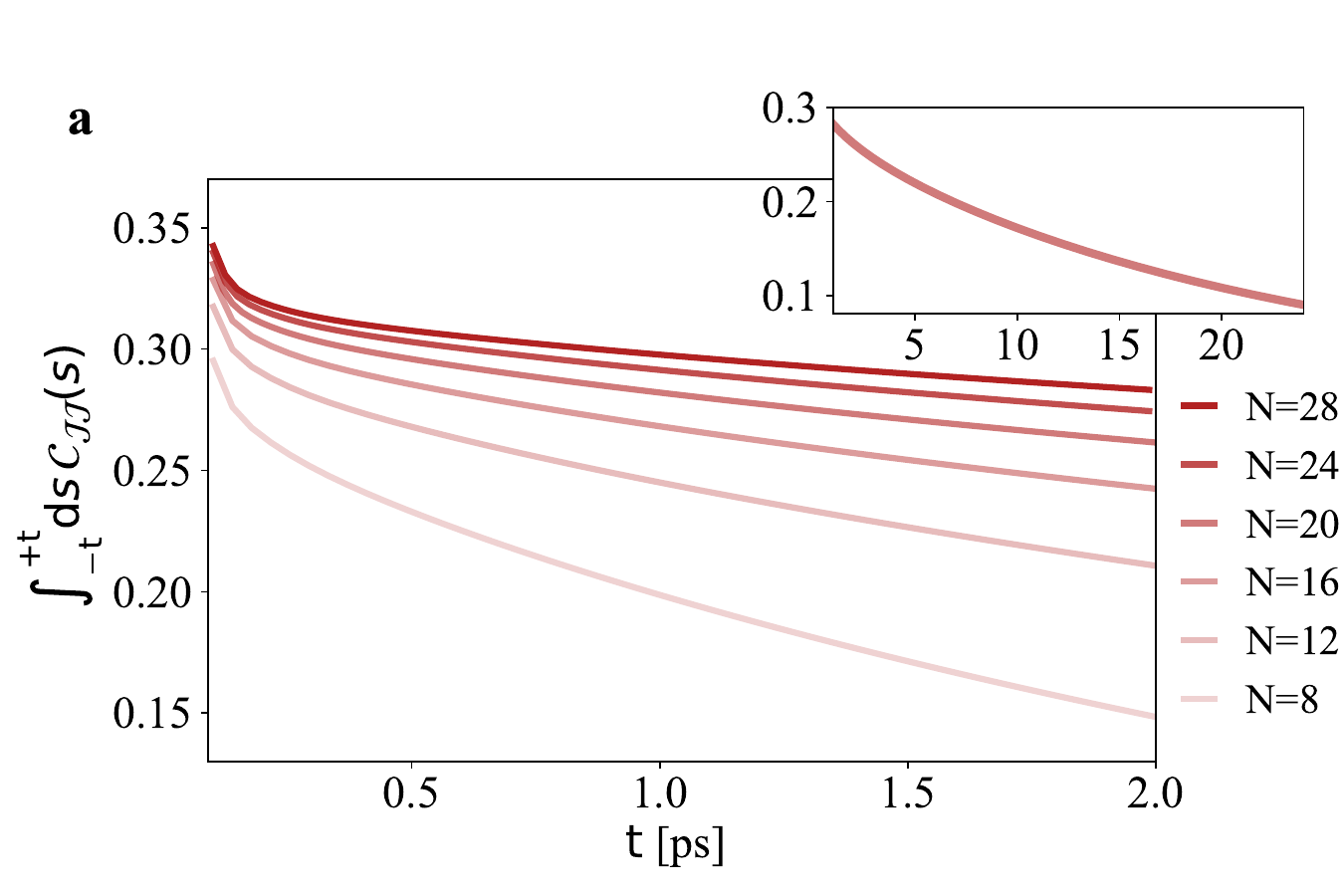}}
    \hspace{10pt}
    \resizebox{.45\textwidth}{!}
    {\includegraphics{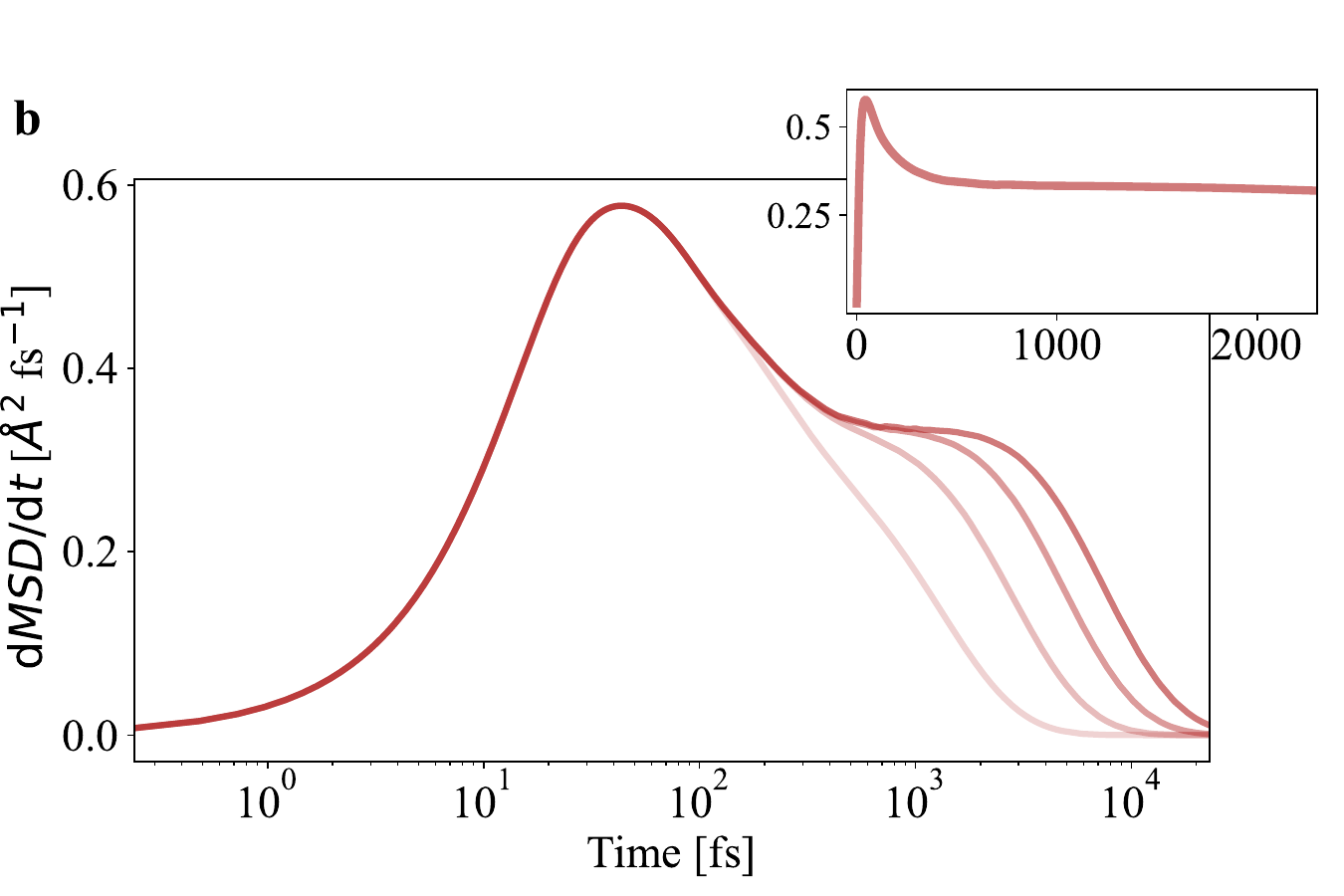}}
\vspace{-20pt}
\end{center}
\caption{\label{fig1} HEOM dynamics performed with $ dt = 0.25$ fs, $\eta = 323$ cm$^{-1}$, $\gamma = 41$ cm$^{-1}$ and $v_{ij} = 50$  cm$^{-1}$ \textbf{a}: Equilibrium simulation: the integral of $C_{JJ}$ against its integration limit for different linear chain lengths. The integral (Eq.~\ref{mu_def_linear_response}) does not converge over reasonable timescales. \textbf{Inset}: Longer simulation for 20-site lattice.
\textbf{b}: Nonequilibrium simulation: time derivative of the MSD against time for different linear chain lengths. The curve for a system size of 20 sites shows a distinct diffusive region, which disappears for shorter chains. For all these studies, we initialize the charge at the middle of the linear chain. \textbf{Inset}: Time derivative of the MSD for 20-site lattice with linear time scale.}
\vspace{-5pt}
\end{figure*}

Theoretical approaches for calculating transport coefficients fall into two major categories: equilibrium and nonequilibrium. Equilibrium approaches invoke linear response theory-derived Green-Kubo relations to encode the nonequilibrium response of the system through the correlation of its equilibrium fluctuations~\cite{kubo1957statistical}, while nonequilibrium methods prepare systems in perturbed states that mirror the desired relaxation process. Regardless of the approach, computing transport coefficients is challenging. A persistent difficulty is the fact that simulations rely on finite systems, which introduce unwanted contributions that only vanish in the macroscopic limit~\cite{steinigeweg2014spin, zhang2004finite, grasselli2022investigating} and misidentifying the influence of the system size has been shown to lead to misdiagnosis of the nature of the transport~\cite{vznidarivc2016diffusive}. Despite the need to account for finite-size effects, differences in their severity between equilibrium and nonequilibrium simulations (in a given system and parameter regime) remain unknown. Yet, the two protocols consider qualitatively different densities in real space (e.g., of matter, charge, or energy), and the idea that the fluxes that they engender experience the system's finite extent equivalently is dubious. 

In the following, we compare the equilibrium and nonequilibrium frameworks for computing transport coefficients. Because these simulations cannot be performed on arbitrarily large systems due to their prohibitive cost, we interrogate a particular example of this problem: the electronic conductivity of the dispersive Holstein model~\cite{holstein1959studies}---a minimal model for conductivity in materials ranging from organic crystals and disordered polymers~\cite{ortmann2010charge}, to covalent organic frameworks~\cite{ghosh2021topology}, and nanomaterials~\cite{mousavi2010electron, cline2018nature}. For example, ionizable functional groups in a polymer, such as unsaturated carbon atoms, are mapped onto a lattice of coupled sites that host mobile electronic excitations (excitons or charges). These excitations couple to, and therefore shift the equilibrium of, a site's local vibrational (bosonic) environment, which spans a broad frequency range characteristic of condensed phase systems~\cite{weiss2012quantum}. This model describes the stabilization of carriers by lattice distortions, resulting in \textit{polaron formation}~\cite{holstein1959studies,mahan2000many, alexandrov2008polarons}. 

While the quantum dynamics of the dispersive Holstein model admit exact numerical solution~\cite{tanimura1989time, suess2014hierarchy}, only relatively short simulation times have been reached. Leveraging algorithmic advances that have become available only in recent years~\cite{shi2009efficient, song2015time}, we have achieved unprecedented simulations in their combination of simulation times ($\sim 25$~ps) and system sizes (N $\leq 40$). These data allow us to capture the physics of polaron formation, determine the timescale of its phenomenology, and disentangle its influence on the transport it exhibits. We show that although both equilibrium and nonequilibrium approaches are expected to yield the same results, these are sensitive to system size in different ways. Further, in certain parameter regimes and model topologies, the nonequilibrium method exhibits diffusion while the equilibrium approach fails to converge at all. 

What is the justification for the equivalence of equilibrium and nonequilibrium approaches? Since the 1950s, researchers have relied on linear response theory, and especially the Green-Kubo relations, to compute transport coefficients spanning thermal conductivities, diffusion constants, magnetic susceptibilities, and viscosity. These relations, built upon Onsager's regression hypothesis, connect the nonequilibrium response of the system to an applied field and the system's equilibrium fluctuations. In our equilibrium protocol, we apply linear response to determine the charge carrier mobility,
\begin{equation}\label{mu_def_linear_response}
    \mu = \frac{1}{2\mathrm{e}_0 \mathrm{k_B} T} \int_{-\infty}^{\infty} \mathrm{d}s \: C_{JJ} (s).
\end{equation}
Here, $\mathrm{e}_0$ is the electron charge, $\mathrm{k_B}$ is the Boltzmann constant, $T$ is the temperature, and $C_{JJ}(t)$ is the quantum current-autocorrelation function at thermal equilibrium.

\begin{figure*}[!tb]
\begin{center} 
\vspace{-40pt}
    \resizebox{.45\textwidth}{!}{\includegraphics{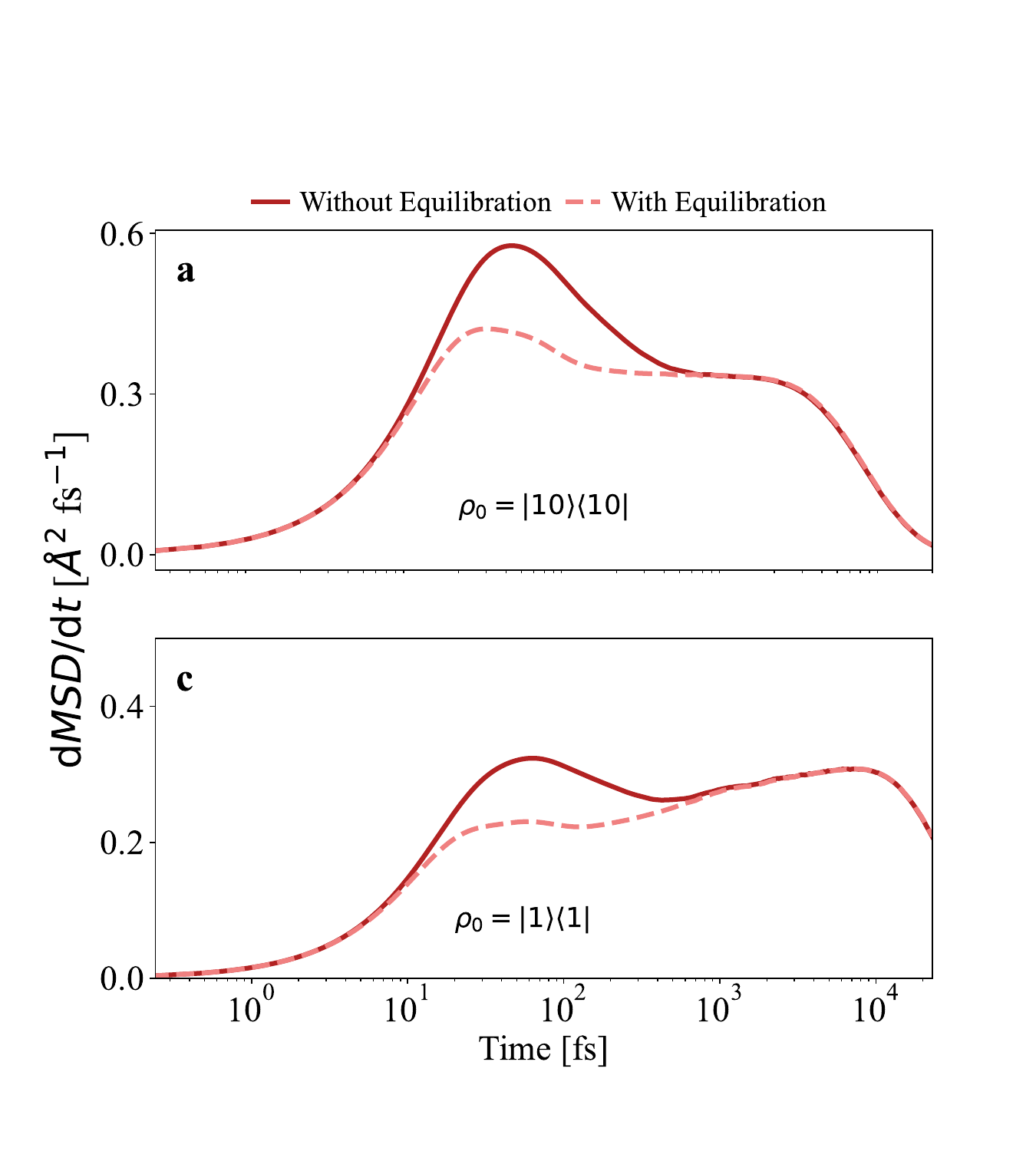}}
    \hspace{-10pt}
    \resizebox{.45\textwidth}{!}
    {\includegraphics{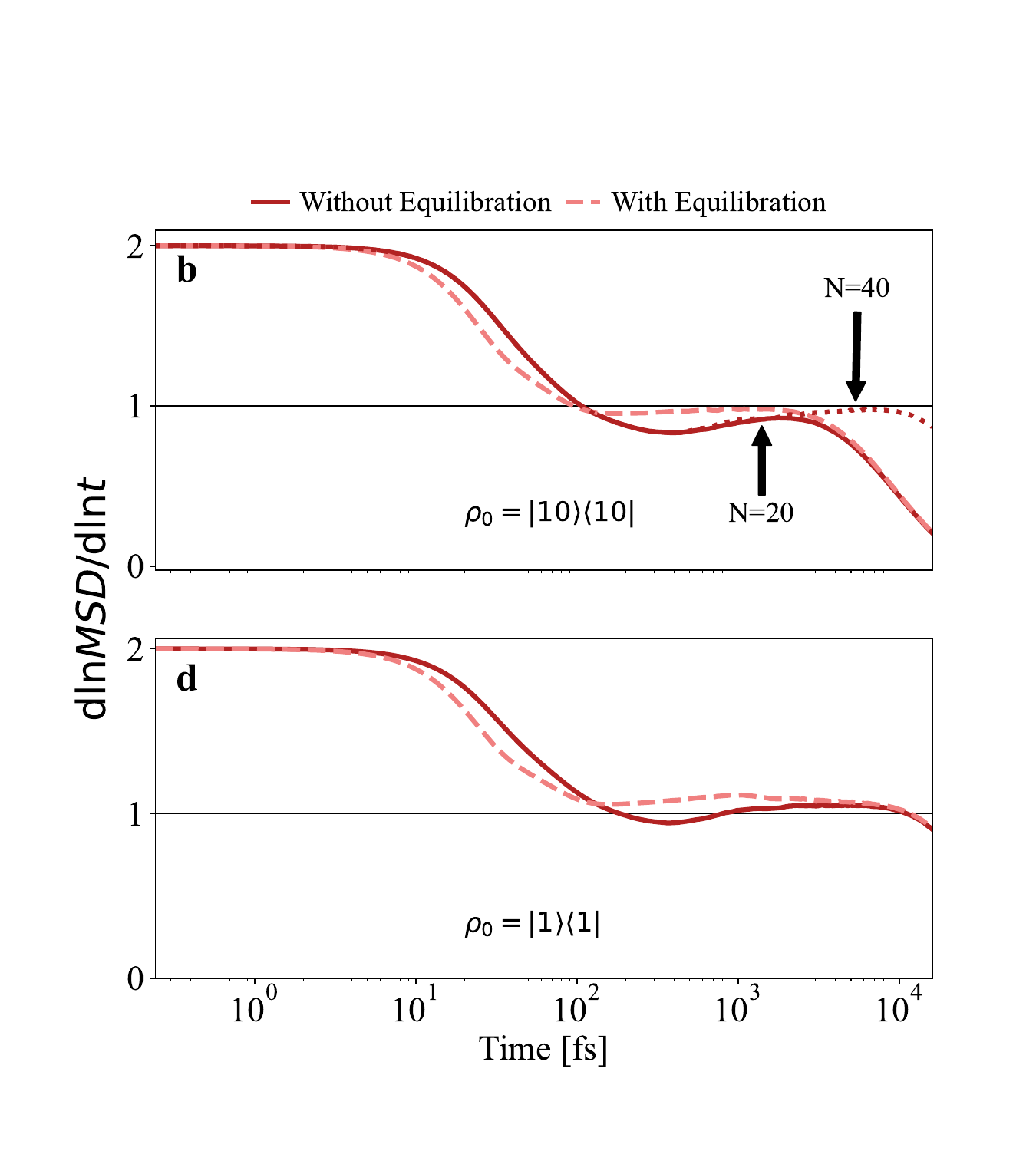}}
\vspace{-30pt}
\end{center}
\caption{\label{fig2} Nonequilibrium HEOM results for a 20-site linear chain. These calculations were performed with $dt = 0.25$~fs, $\eta = 323$~cm$^{-1}$, $\gamma = 41$~cm$^{-1}$, and $v_{ij} = 50$~cm$^{-1}$. \textbf{a}: $\rmm{d}MSD/\rmm{d}t$ when the charge is initialized at the center of the linear chain. We can observe an approximately flat region $\sim$ 600--1600~fs. \textbf{b}: Corresponding plot for $\alpha$, which shows the initial ballistic region, followed by a subdiffusion towards a steady value of the $\alpha$ before the finite-size effect. The dotted line is the same simulation but for N=$40$ \textbf{c}: $\rmm{d}MSD/\rmm{d}t$ when the charge is initialized next to one wall. In this case, the derivative never plateaus. \textbf{d}: Corresponding plot for $\alpha$ showing a similar initial ballistic region, followed by a superdiffusive region instead.}
\vspace{-5pt}
\end{figure*}

In contrast, the widely adopted nonequilibrium protocol for determining charge mobility employs the Einstein relation, $\mu = \mathrm{e}_0D/\mathrm{k_B} T$, to connect $\mu$ to the diffusion constant $D$ of a charge (see the Methods section for details). At early times the mean square displacement (MSD) can display complex dynamics. Once the motion transitions from ballistic to diffusive transport, the MSD exhibits linear growth in time. In an infinite system, diffusion continues indefinitely, inspiring the definition
\begin{equation}\label{def-msd}
    D = \lim_{t \to \infty} \frac{1}{2}\frac{\mathrm{d}MSD (t)} {\mathrm{d}t}.
\end{equation}

We begin our analysis by performing a series of equilibrium and nonequilibrium simulations for the dispersive Holstein model parameterized to embody typical organic semiconductors~\cite{nematiaram2020modeling}. 
We have employed the numerically exact hierarchical equations of motion (HEOM)~\cite{tanimura1989time} to obtain the dynamics. In the equilibrium protocol, one must evaluate the integral of $C_{JJ}(t)$ over infinite limits (Eq.~\ref{mu_def_linear_response})~\cite{song2015new}. At long times, this correlation function approaches zero since $\lim_{t\rightarrow \infty}C_{JJ}(t) = \langle \hat{J} \rangle^2$ and $\langle \hat{J} \rangle=0$. We define the timescale at which this occurs as the equilibrium time, $t_\mathrm{eq}$. Consequently, we compute the integral as a function of the limits, anticipating that the curve plateaus at the value $2\mathrm{e}_0\mathrm{k}_\mathrm{B}T\mu$.

Yet, Fig.~\ref{fig1}\textbf{a} demonstrates that, for an open chain, the integral in Eq.~\ref{mu_def_linear_response} does not converge with $t$. Examining the correlation function (Fig.~S1) reveals that $C_{JJ}$ approaches zero from below, converging faster for longer chains. This reveals the presence of a finite-size effect. Although our simulations reach unprecedented system sizes and simulation times, the parameters required to converge this integral remain unaffordable. The failure to return a meaningful value for $\mu$ suggests an anomalous transport beyond the scope of linear response.

What kind of transport does the nonequilibrium treatment reveal for this same system? Figure~\ref{fig1}\textbf{b} reveals that the protocol exhibits a well-behaved diffusion region for an open chain of~$N \geq 20$~sites. Thus, the nonequilibrium method converges with increasing system size without any hint of persistent anomalous transport. This striking difference---that one method can circumvent finite-size effects fatal to the other---proves the two approaches interact with the system size in qualitatively different ways and motivates a deeper analysis.

Unlike the equilibrium case, the nonequilibrium protocol is sensitive to the initial state of the system. Hence, we place the charge on a single site of the chain and test the sensitivity of the results to the extent of local lattice equilibration around the charge prior to its migration. We test two extremes for the initial state preparation:
\begin{enumerate}
    \item[1.] With equilibration (w/ eq.)---the charge is allowed to completely equilibrate with its local vibrational environment on a particular site. This corresponds to a fully localized polaron state. The charge migration starts after local equilibration is complete. 
    
    \item[2.] Without equilibration (w/o eq.)---the charge is placed on a lattice site and released before its local vibrational environment has reacted. Polaron formation and charge migration begin simultaneously. 
\end{enumerate}

These two different initializations produce different early-time derivatives of the MSD, evident in the difference between the full and dashed lines before $\sim\!\!550$~fs in Fig.~\ref{fig2}. If the system has a well-defined diffusion constant, these two curves must exhibit the same (flat) slope at later times. Therefore we can divide the dynamics into four distinct nonequilibrium regimes. 

The first, ballistic regime quickly transitions to a period of semi-local polaron equilibration that ends when the two traces meet at $\tau_P\!\sim\!550$~fs. We qualitatively identify the third regime as diffusive motion, beginning immediately after $\tau_P$. In an infinite system, this plateau would continue indefinitely as described by Eq.~\ref{def-msd}. However, here, the system enters the fourth regime beginning $\sim 1600$~fs and the curve decreases to zero. This time, which we call $\tau_R$, corresponds to when the charge reaches the boundary of the system, reflects back, and interferes with itself. Thus the decrease in $\mathrm{d}MSD/\mathrm{d}t$ toward zero signals the emergence of the finite-size effect for this method (see Fig.~S2). It follows that the success of the nonequilibrium simulation relies on the separation in timescales between the reflection time $\tau_R$ and the polaron relaxation time $\tau_P$. 

\begin{figure}[!b]
\vspace{-8pt}
\begin{center} 
    \resizebox{.5\textwidth}{!}{\includegraphics[trim={0 0 0 60pt},clip]{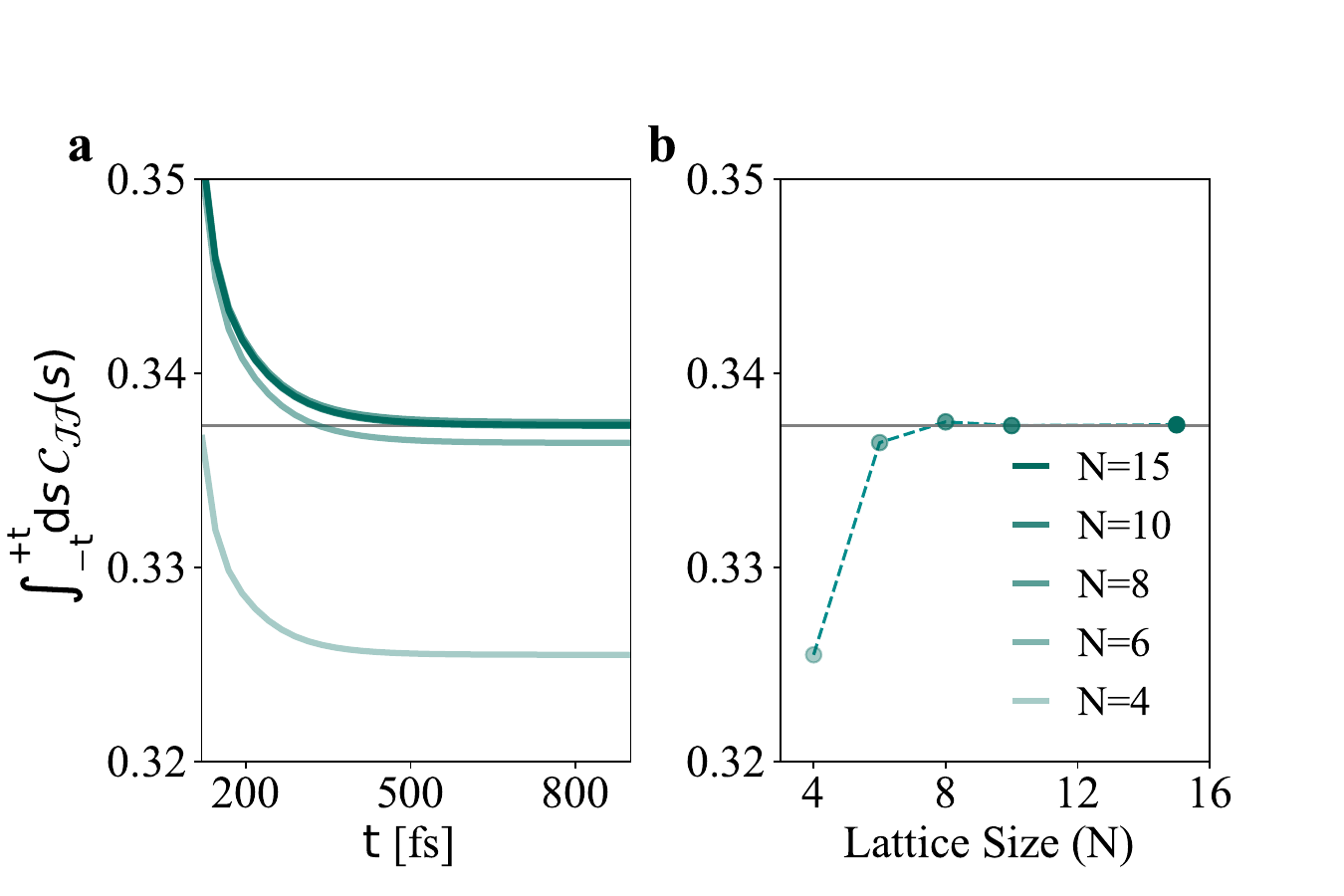}}
\vspace{-20pt}
\caption{\label{fig3} Equilibrium HEOM results for the periodic topology. The dynamics were calculated with $dt = 0.25$~fs, $v_{ij} = 50$~cm$^{-1}$, $\eta = 323$~cm$^{-1}$, and $\gamma= 41$~cm$^{-1}$. \textbf{a}: Integral of $C_{JJ}$ against its integration limit for different periodic chain lengths. The integral converges both with $t_\mathrm{eq}$ and system size. \textbf{b}: Convergence plot of the integral against ring size (final point of the left panel). The grey dashed line in both panels shows the final, converged value as being 0.337~\AA$^2$fs$^{-1}$.}
\vspace{-14pt}
\end{center}
\end{figure}

The semi-local equilibration of the polaron determining $\tau_P$ can be understood as the decreasing difference in on-site charge density for the w/ eq.~and w/o eq.~cases as a function of both time elapsed and distance from the initial site (see Fig.~S3)~\cite{golevz2012relaxation}. In other words, both nonequilibrium simulations converge to the same charge distribution after an initial period of polaron relaxation. We quantify this via the scaling exponent of the $MSD \propto |t|^\alpha$ against (log) time. As Fig.~\ref{fig2}\textbf{b} shows, $\alpha = 2$ for ballistic expansion at early times, but quickly decreases to $\alpha = 1$, signaling diffusion. The region of the curve between these two limits holds essential information about the nature of $\tau_P$. Specifically, the relatively larger value of $\alpha$ for the w/o eq.~case during the first $\sim 100$~fs reports on the initially lower drag compared to the w/ eq.~case, where a fully formed polaron presents an activation barrier to charge transfer. Further, careful inspection of $\alpha$ reveals that the w/o eq.~simulation does not actually reach the diffusive regime before the onset of finite-size effects. If our argument is correct, we should expect to see convergence to $\alpha = 1$ from below if we perform the (extremely costly) N$=40$ simulation to increase the time of unadulterated dynamics. Indeed, the dotted line in Fig.~\ref{fig2}\textbf{b} confirms our prediction. Moreover, even the w/ eq.~simulation visits the region $\alpha < 1$, which offers hints at the relative contribution from polaron formation on different sites. Physically, the formation of polarons on sites increasingly far from the original site corresponds to an increasing drag experienced by the charge carriers as they migrate along the chain, which dynamically manifests as `subdiffusion' or $\alpha < 1$.

Confirmation of $\tau_R$ as the reflection time is simpler, as we can numerically ascertain the time at which the charge first reaches the final site.  At some critical system size, this causes $\tau_P > \tau_R$ and the disappearance of the diffusive region. At this point, the finite-size effect becomes fatal for the nonequilibrium protocol, as is evident from the results obtained when reducing the lattice from $N=16$ to $N=8$ in Fig.~\ref{fig1}\textbf{b}. The dependence on system size implies that, even on a long chain, if one were to initiate the charge on a site closer to a boundary, reflection would onset sooner. However, we find the situation becomes murky when considering an extreme case like that depicted in Fig.~\ref{fig2}\textbf{c}. Here, initiating the charge immediately next to the wall leads to a qualitatively different behavior of the MSD, where $\mathrm{d}MSD/\mathrm{d}t$ increases even \textit{after} $\tau_P$, preventing the identification of a diffusive region before $\tau_R$ altogether. This suggests the presence of a persistent asymmetric bias or `wall effect' in the dynamics, which is unrelated to the chain length. Physically, after an initial rise, the carrier density on each site is found to monotonically approach the equilibrium value from above (see Fig.~S4), meaning that a persistent directional bias emerges which precludes the onset of normal diffusion. We can further confirm this by plotting the value of $\alpha$ in Fig.~\ref{fig2}\textbf{d}, which for both cases reaches a stable value of $\sim\!1.05$ before $\tau_R$, i.e. a steady-state with superdiffusive character. As a result, this simulation \textit{never} exhibits typical transport, even when $N$~is increased many times.

Given that the proximity of the charge to a wall can prevent the nonequilibrium simulations from capturing diffusive behavior, one may ask if a similar effect could cause the lack of convergence in the equilibrium approach. After all, the initial condition of the system in a linear response method is the canonical equilibrium density, $\propto e^{-\beta \hat{H}}$, which places charge on every site. Therefore, regardless of the chain length, the equilibrium calculation always contains a component with knowledge of the wall, suggesting that the failure to determine the transport coefficient is unique to the open chain \textit{topology}. Recent simulations have noted a similar lack of convergence in such topologies~\cite{bertini2021finite}. Hence, to test our hypothesis we turn to equilibrium simulations under periodic boundary conditions.

Figure~\ref{fig3} presents the results of equilibrium simulations for a periodic chain with all parameters being equal to those of the linear case analyzed above. As the chain length increases from $N=4$, we obtain rapid convergence to the $N=15$ value of 0.337~\AA$^2$~fs$^{-1}$ as early as $N=8$. In all cases, the integral in Eq.~\ref{mu_def_linear_response} converges fully by 700~fs, in stark contrast to the linear case of Fig~\ref{fig1}\textbf{a}, which failed to converge even after $25$~ps. This confirms the validity of our hypothesis and demonstrates that the failure of the equilibrium simulation is dependent on the linear chain topology. This \textit{wall effect} has a profound implication: since knowledge of the wall is independent of chain length, fully eliminating the finite-size effect in the equilibrium dynamics of open chains requires an infinitely large system. This has deep implications for the use of methods that employ tensor network decompositions, which perform most efficiently in open-chain topologies~\cite{strathearn2018efficient}.

Throughout this analysis, we employed a parameter regime in which the nonequilibrium simulation clearly exhibits a diffusive region. However, timescale competition between $\tau_P$ and $\tau_R$ can cause a breakdown in the nonequilibrium protocol where it is impossible to identify a plateau in $\rmm{d}MSD/\rmm{d}t$, as in the small-$N$ traces of Fig.~\ref{fig1}\textbf{b}. Understanding when the separation of these timescales allows the diffusion region to emerge and, therefore, yields the transport coefficient determines the minimum $N$ required to eliminate the finite-size effect. We thus turn to a thorough interrogation of the physically relevant parameter space.

\begin{figure}[!ht]
\begin{center} 
\vspace{2pt}
    \resizebox{.5\textwidth}{!}{\includegraphics[trim={0 0 0 30pt},clip]{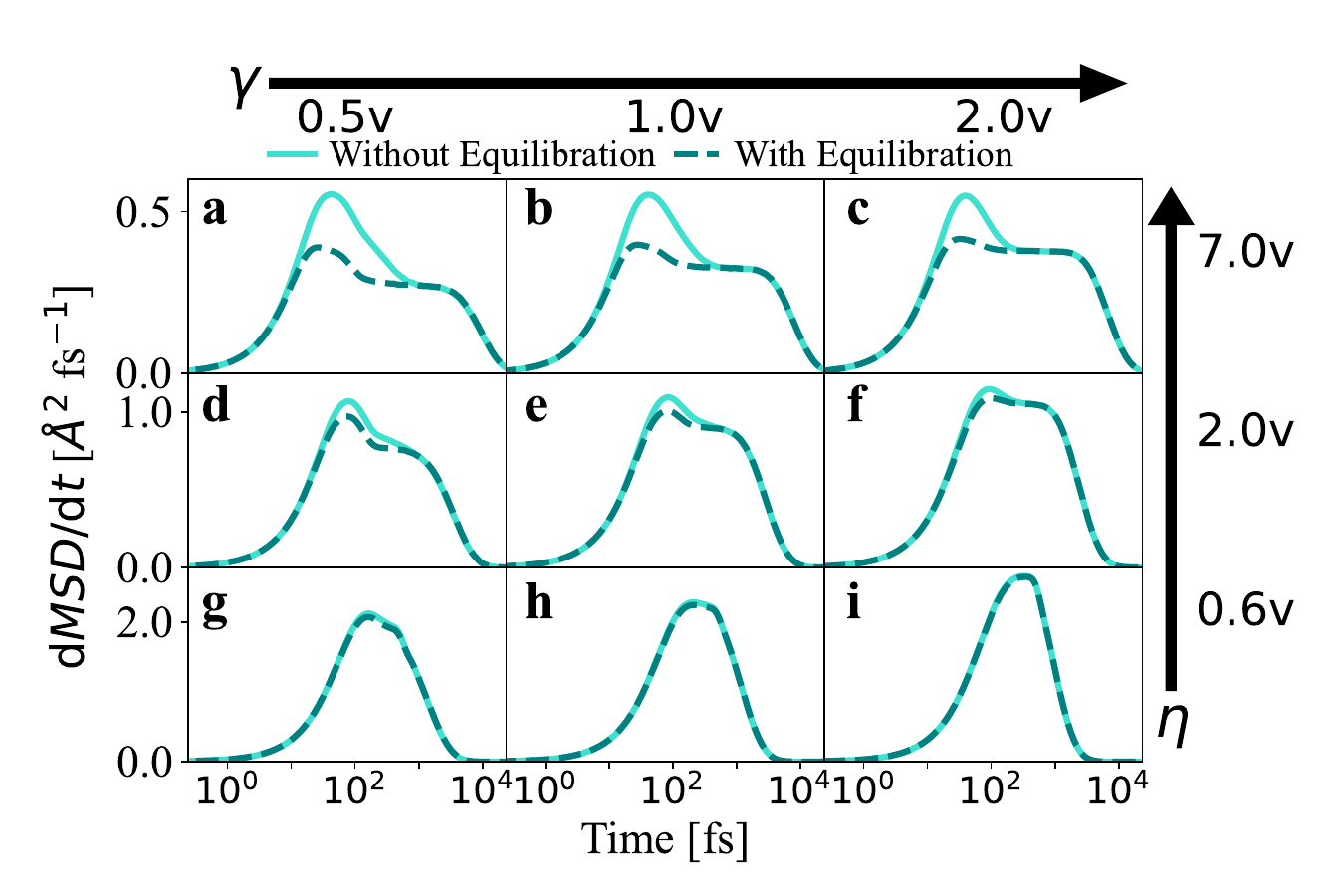}}
\vspace{-20pt}
\caption{\label{fig4} Nonequilibrium HEOM results for a 20-site periodic chain. These calculations were performed with $dt = 0.25$~fs and $v_{ij} = 50$~cm$^{-1}$. From top to bottom: Decreasing reorganization energy (i.e., charge-phonon interaction strength) which manifests as an earlier onset of reflection timescale $\tau_R$ and decrease of the polaron relaxation effect's magnitude (but not $\tau_P$ itself). From left to right: Increasing the bath speed which reduces the initial polaron effect's signature (both relaxation time $\tau_P$ and magnitude)}
\vspace{-25pt}
\end{center}
\end{figure}

Figure~\ref{fig4} shows the nonequilibrium dynamics for a periodic Holstein chain as a function of two parameters: increasing charge-phonon coupling in the form of the reorganization energy (from bottom to top) and increasing phonon-environment decorrelation speed (from left to right). The results recapitulate the four nonequilibrium regimes: ballistic transport, semi-local polaron relaxation, the putative diffusion regime for which $\alpha \rightarrow 1$ at long times (when it appears), and the reflection time that emerges as a consequence of finite system sizes.  Inspection of Fig.~\ref{fig4} reveals that:

\begin{itemize}
    \item While a homogeneous periodic lattice eliminates the issue of choosing the initialization site from the nonequilibrium perspective, one still observes the downturn in $\rmm{d}MSD/\rmm{d}t$, which arises from the charge interfering with itself when it reaches the diametrically opposite point in the ring. Thus, periodic lattices also exhibit a competition of timescales that manifests as a finite-size effect. 
    \item As the phonon environment speed increases (going across a row), the duration of the semi-local polaron relaxation shortens, leading to a faster onset of diffusive motion and a larger diffusion constant. 
    \item Diminishing the reorganization energy (going down a column) reduces the semi-local polaron relaxation. Additionally, the reflection regime onsets earlier, consistent with the barrier to hopping lowering. While less evident, the onset of the diffusion region ($\tau_P$) remains unchanged. 
    \item As the reorganization energy becomes smaller than the intersite coupling (bottom row), the diffusive region disappears, as the polaron relaxation and reflection timescales overlap. Capturing the diffusion region in this parameter regime requires significantly larger lattices.
\end{itemize}

These results demonstrate that the severity of finite-size effects depends strongly on the parameter regime. To observe diffusion when $\gamma = 0.5 v$ and $\eta = 0.6 v$ (Fig.~\ref{fig4}\textbf{g}) one needs a larger system where $\tau_R > \tau_P$. This begs the question: does an equivalent size requirement emerge in the equilibrium approach? That is, do weakly coupled systems require bigger rings for equilibrium calculations to converge, and how does this value of $N$ compare to the nonequilibrium case? We find that the equilibrium approach, for this choice of parameters, converges by $N=20$ (see Fig.~S5), whereas the nonequilibrium simulations for $N=20$ (Fig.~\ref{fig4}\textbf{g}) fail to reveal a plateau region. What is more, even in the far easier parameter regime of Fig.~\ref{fig2} (Fig.~\ref{fig4}\textbf{b}), the nonequilibrium simulation requires around $N=40$ to reach truly diffusive behavior, and therefore the correct value of $\rmm{d}MSD/\rmm{d}t$. This competition of timescales in nonequilibrium simulations and, in particular, the system size-dependent reflection time, means that nonequilibrium simulations require large system sizes to reveal the nature of transport and a converged transport coefficient. In contrast, the equilibrium approach, in this periodic topology where it functions correctly, is considerably cheaper to converge.

Our results thus enable us to determine the source of anomalous transport in small polaron-forming systems: semi-local nonequilibrium lattice relaxation, over a broad range of timescales, that generates spatial heterogeneity. Prior to complete lattice equilibration, which occurs long after $\tau_R$, each site encounters varying charges. This transient variability results in different extents of polaron formation across the lattice, akin to the heterogeneity observed in biomolecules within the cytoplasm or particles in crowded fluids. This leads to anomalous transport. Only close to equilibrium, when the charge discrepancy between neighboring sites diminishes considerably, does heterogeneity become vanishingly small. For this reason, equilibrium simulations easily capture the transport coefficient (indicating normal transport) when implemented in a periodic system. Nevertheless, in most parameter regimes, one can afford sufficiently large system sizes, yielding a nonequilibrium value for $\alpha$ which is close to one. Such $\alpha$ values enable one to adequately determine the transport coefficient in the nonequilibrium case.

We anticipate that our work, especially when combined with advances in electronic structure theory~\cite{giustino2017electron}, will offer a unifying framework to understand and control small polarons across experiments. Indeed, cutting-edge measurements in systems spanning perovskites~\cite{delor2020imaging}, conjugated polymers~\cite{penwell2017resolving}, nanomaterials~\cite{akselrod2014subdiffusive}, and transition metal oxide photocatalysts~\cite{carneiro2017excitation} bolster the idea that polaron dynamics stands to guide materials design for optimized transport properties. 

\clearpage

\section*{Methods}

\renewcommand{\theequation}{M.\arabic{equation}}
\setcounter{equation}{0}

\vspace{-6pt}
\subsection{Hamiltonian}
\vspace{-6pt}

The Holstein Hamiltonian has been widely used to capture the physics of charge transport phenomena in condensed phase systems. While the original Holstein model assumes local coupling to a \textit{single} optical phonon mode~\cite{holstein1959studies,mahan2000many, alexandrov2008polarons}, we study the dispersive Holstein model, which is more appropriate for organic crystals and disordered polymers~\cite{nematiaram2020modeling},
\begin{subequations}
\begin{align}
    \hat{H} &= \sum_{i}^{N}\epsilon_i \hat{a}_i^\dag \hat{a}_i + \sum_{\langle ij \rangle }^{N} v_{ij} \hat{a}_i^\dag \hat{a}_j \nonumber \\
    &\quad +  \sum_{i}^{N}\big[ \hat{H}_{B, i} + \hat{V}_{B, i}\hat{a}_i^\dag \hat{a}_i \big], \label{ham-eq} \\
    \hat{H}_{B,i} &= \frac{1}{2}\sum_{\alpha} \big[ \hat{P}_{i,\alpha}^2 +  \omega_{i\alpha}^2 \hat{X}_{i,\alpha}^2 \big], \label{eq:free-bath-ham}\\
    \hat{V}_{B,i} &= \sum_{\alpha} c_{i, \alpha} \hat{X}_{i,\alpha} \label{eq:linear-coupling}.
\end{align}
\end{subequations}
Here $\hat{a}_i$ ($\hat{a}_i^\dag$) is the annihilation (creation) operator for an electron/hole on lattice site $i$ and $\hat{X}_{i,\alpha}$ ($\hat{P}_{i,\alpha}$) is the dimensionless position (momentum) operator for $\alpha$-th phonon mode connected with $i$-th lattice site. The total number of sites is $N$ and each site is connected to a bath of phonon modes. The first term of Eq.~\ref{ham-eq} sums over the energy contributions of occupied sites and the second term allows charges to hop from one site to its nearest neighbors (denoted by the angular brackets). The third term (explicitly given in Eq.~\ref{eq:free-bath-ham}) accounts for the free vibrational Hamiltonian, and the final term contains the on-site electron-phonon coupling, which is linear in the local phonon bath coordinates (see Eq.~\ref{eq:linear-coupling}). This coupling is responsible for the fluctuation of the on-site energies and ultimately results in polaron formation, where the lattice distorts around a charge offering additional stabilization to that particular (localized) configuration.

The charge-phonon interaction through which the $i$-th  site is connected with its phonon modes is described by the spectral density,
\begin{equation}\label{spec_dens_def1}
   J_i(\omega)= \dfrac{\pi }{2} \sum_{\alpha} \dfrac{c_{i,\alpha}^2} {\omega_{i\alpha}} \delta(\omega -\omega_{i\alpha}).
\end{equation}
We assume that all $J_i (\omega)$ are equivalent for all sites and follow a Debye form commonly used to capture the dissipation in the condensed phase~\cite{weiss2012quantum} 
\begin{equation}\label{spec_dens_def2}
    J(\omega)= \dfrac{ \eta \gamma \omega}{\omega^2 + \gamma^2},
\end{equation}
where $\eta/2$ is the reorganization energy and $1/\gamma$ is the timescale at which the phonon environment decorrelates. Aligning with previous studies~\cite{song2015new}, we set $\epsilon_i$ to 0. The lattice distance between two consecutive sites is chosen to be $a = 5~$\AA~without loss of generality. We also choose all the nearest neighbor coupling to be the same, i.e., $v_{ij} =v$. 

\vspace{-12pt}
\subsection{Current Autocorrelation Function}
\vspace{-6pt}

The current autocorrelation function $C_{JJ}$ is defined as 
\begin{equation}\label{cjj-def}
    C_{JJ} = \frac{1}{Z}\text{Tr}\big[e^{-\beta \hat{H}} \hat{J}(t) \hat{J}(0)  \big],
\end{equation}
where $\hat{H}$ is the full Hamiltonian of the system, $Z = \mathrm{Tr}[e^{-\beta \hat{H}}]$ is the partition function of the total system, $\hat{J}(t)$ is the current operator at time $t$, and $\beta=1/\mathrm{k}_\mathrm{B}T $. The current operator is defined as~\cite{mahan2000many}
\begin{equation}
    \hat{J} = -\frac{ia\mathrm{e}_0} {\hbar} \sum_{\langle mn \rangle} v_{mn} (\hat{a}_m^\dag \hat{a}_n -\hat{a}_n^\dag \hat{a}_m),
\end{equation}
where $i = \sqrt{-1}$ and $\hbar$ is Planck's constant. In all results shown, we have set $\mathrm{e}_0 = 1$.

\vspace{-6pt}
\subsection{Mean Squared Displacement}
\vspace{-6pt}

We initialize the charge at a particular site of the lattice and observe the population at different sites as a function of time. One can compute its mean-squared displacement, 
\begin{equation}\label{msd-def}
MSD(t) = \sum_{i} d_i^2 P_i (t), 
\end{equation}
where $ d_i$ represents the distance between the $i$th site and the initial lattice site, and $P_i(t) = \mathrm{Tr}[\hat{\rho}(t) \hat{a}_i^{\dagger}\hat{a}_i]$ is the charge population at the $i$th site at time~$t$. 

\vspace{-12pt}
\subsection{Simulation Details}
\vspace{-6pt}

We have employed the hierarchical equations of motion (HEOM) to obtain the numerically exact dynamics of the dispersive Holstein system at finite temperatures~\cite{tanimura1989time}. HEOM maps the entire open quantum system into auxiliary reduced density matrices (ADMs) and solves coupled differential equations for all the ADMs at each time step. We converged the HEOM calculation with respect to the internal parameters of the method: hierarchical depth $L=26$ and Matsubara parameter $K=2$. However, these calculations are prohibitively expensive both in memory and simulation time. Therefore, to render them affordable, we have employed dynamical filtering and an advanced $n$-particle approximation that allows us to retain the numerical exactness of our result while minimizing the cost~~\cite{shi2009efficient, song2015time}. To ensure consistency with previous works, we have performed all the simulations in the dilute electron/hole limit where there is only one charge in the system.

To compute the nonequilibrium relaxation simulation, we initialize the charge at the $i$th site by making the $i$th diagonal element of the reduced density matrix (RDM) equal to unity. To implement the w/ eq.~strategy, we initiate the charge, set the nearest neighbor hopping term $v_{ij}=0$, and propagate all ADMs using HEOM until the density stops changing. This allows for local polaron formation on the $i$th site, yielding $\hat{\rho}(t\rightarrow \infty) = \hat{a}_i^{\dagger}\hat{a}_i \hat{\rho}_{B,i}^{\rm shifted} \prod_{k \neq i} \hat{\rho}_{B,k}$, where $\hat{\rho}_{B,i}^{\rm shifted} = e^{-\beta (\hat{H}_{B,i} + \hat{V}_{B,i})} / \mathrm{Tr}_{B,i}[e^{-\beta (\hat{H}_{B,i} + \hat{V}_{B,i})}]$ and $\hat{\rho}_{B,k} = e^{-\beta \hat{H}_{B,k} } / \mathrm{Tr}_{B,k}[e^{-\beta 
\hat{H}_{B,k}}]$. After this initial equilibration, we allow the charge to hop by making $v_{ij}$ nonzero and trace the charge density at each site with time. In contrast, the w/o eq. initial condition is simply $\hat{\rho}(0) = \hat{a}_i^{\dagger}\hat{a}_i \prod_{k}  \hat{\rho}_{B,k}$.

We follow the protocol in Ref.~\cite{song2015new} to compute equilibrium correlation functions using HEOM. This calculation is split into two parts. First, we propagate the reduced density matrix of the electronic degrees of freedom and all ADMs for timescales that are sufficiently long to achieve equilibrium. This calculation yields the correlated
equilibrium density, $\hat{\rho}_{eq} = e^{-\beta \hat{H}}/Z$. Then all the ADMs associated with this correlated state are multiplied by the current operator $\hat{J}$ to make a modified initial condition $\hat{\tilde{\rho}}(0) = \hat{J}\hat{\rho}_{eq}$, which we then propagate forward in time. The correlation function is calculated as $C_{JJ} = \text{Tr}[\hat{J}\hat{\tilde{\rho}}(t) ]$. This expression is equivalent to that in Eq.~(\ref{cjj-def}).

The dynamic filtering parameters are
$\delta = 10^{-7}$
for nonequilibrium simulations (both w/ eq.~and w/o eq.~cases) and initial state preparation for equilibrium simulation and $\delta = 10^{-10}$ (in atomic
units) for the post-equilibration propagation in the calculation of equilibrium correlation functions.

\section*{Acknowledgments}

A.M.C. acknowledges the start-up funds from the University of Colorado Boulder for partial support of this research. Acknowledgment is made to the donors of the American Chemical Society Petroleum Research Fund for partial support of this research (No.~PRF 66836-DNI6). We thank Prof.~Qiang Shi for sharing his HEOM code with us. This work utilized the Alpine high-performance computing resource at the University of Colorado Boulder. Alpine is jointly funded by the University of Colorado Boulder, the University of Colorado Anschutz, Colorado State University, and the National Science Foundation (award 2201538).

\bibliography{ref1.bib}

\onecolumngrid
\clearpage
\section*{Supplementary Information}
\twocolumngrid
\renewcommand{\thefigure}{S\arabic{figure}}
\setcounter{figure}{0}
\renewcommand{\theequation}{S.\arabic{equation}}
\setcounter{equation}{0}

\begin{figure}[!ht]
\vspace{-6pt}
\begin{center} 
    \resizebox{.45\textwidth}{!}{\includegraphics{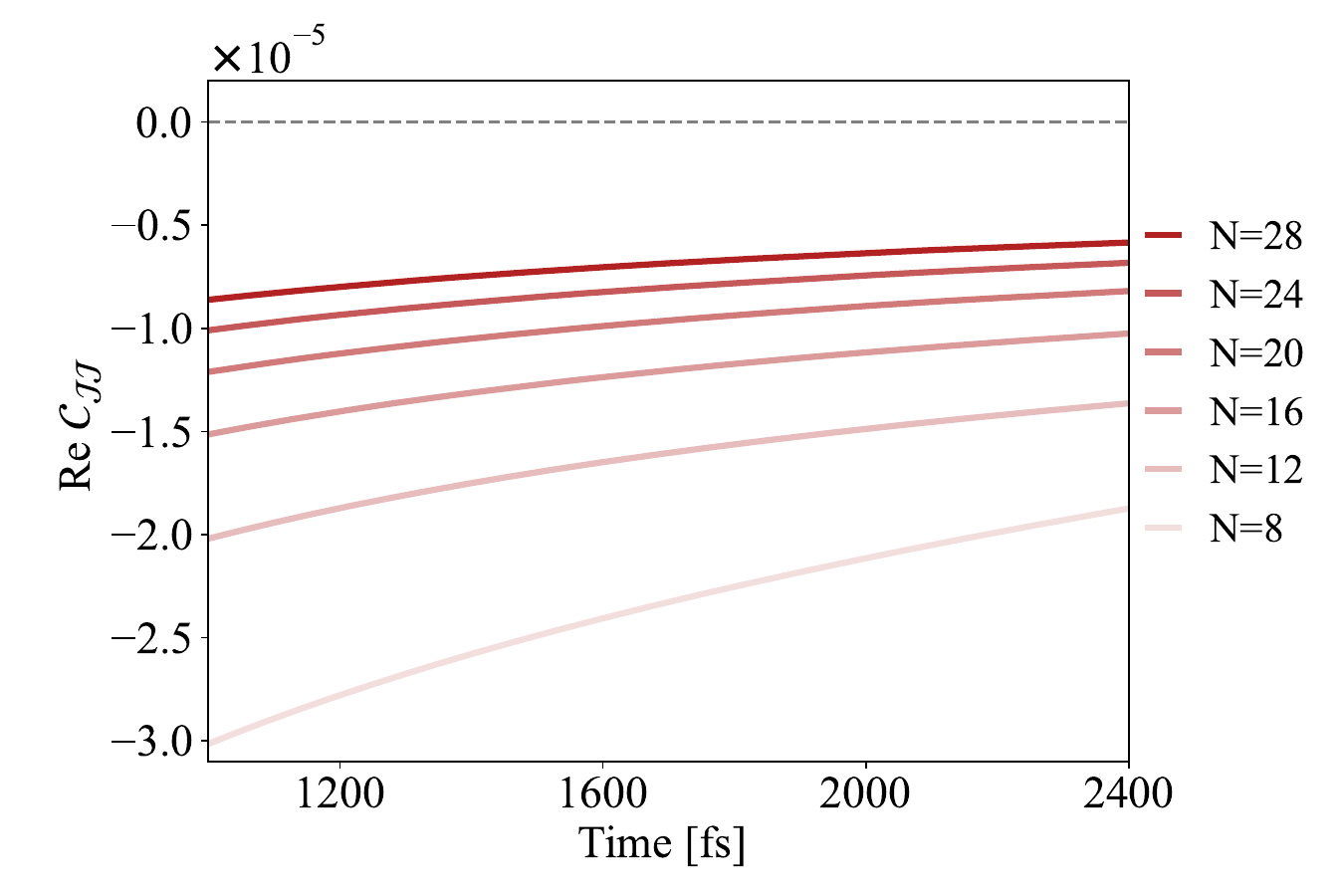}}
\vspace{-12pt}
\caption{\label{fig:real_part_current} HEOM calculations were performed with $ dt = 0.25$ fs, $\eta = 323$ cm$^{-1}$, $\gamma = 41$ cm$^{-1}$ and $v_{ij} = 50$  cm$^{-1}$ to compute the current autocorrelation function for different system size $N$. Plot displays how the real part of current autocorrelation approaches towards zeros as we increase $N$.}
\end{center}
\end{figure}

\begin{figure}[!ht]
\vspace{-6pt}
\begin{center}
    \resizebox{.5\textwidth}{!}{\includegraphics{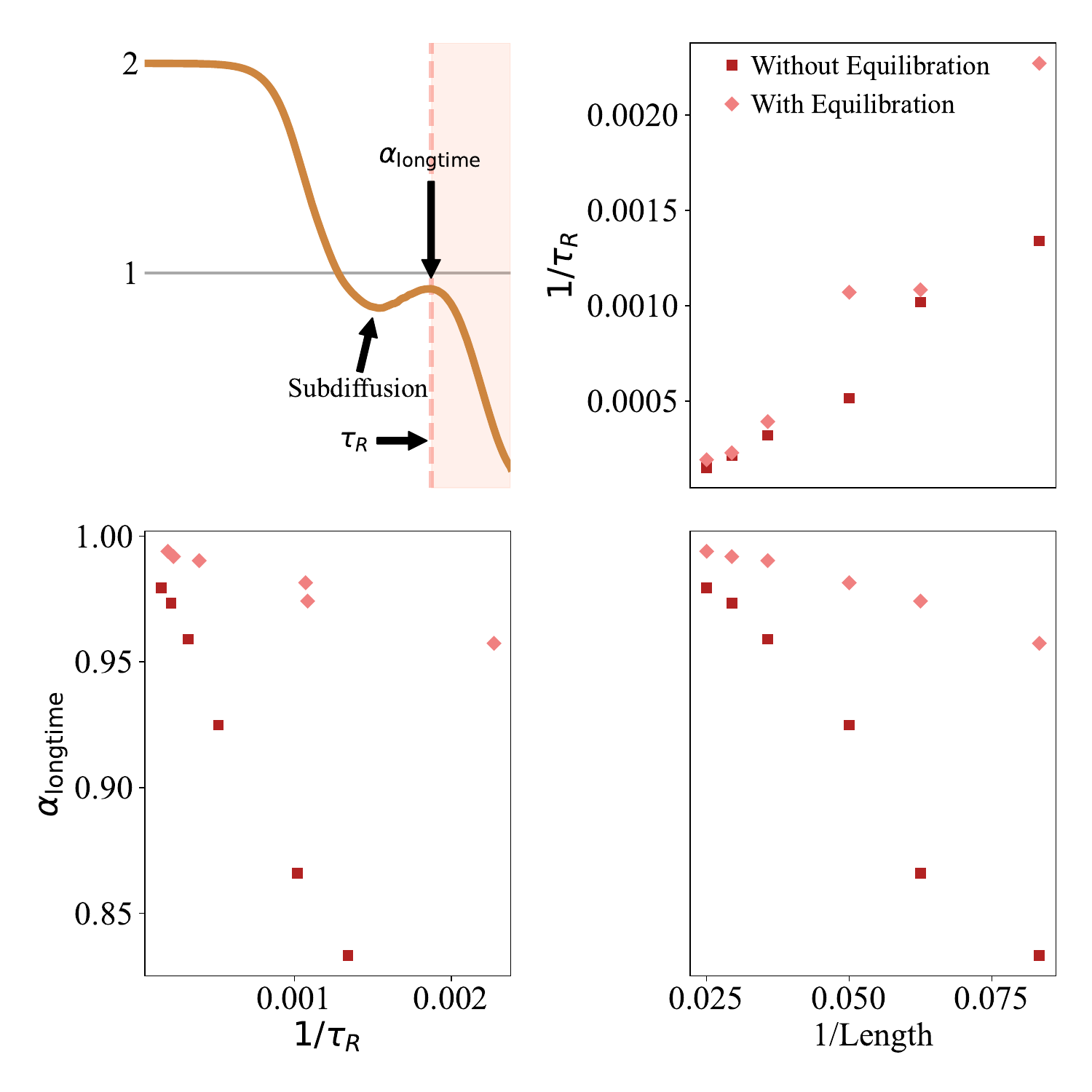}}
\vspace{-24pt}
\caption{\label{fig:alpha_all} \textbf{Top Left}: Cartoon showing how $\alpha$ changes from ballistic to subdiffusive, and then exhibits a plateau region. The corresponding maximum value $\alpha$ is chosen as $\alpha_{\rm{longtime}}$ and the time is defined as $\tau_R$, after this $\alpha$ falls due to reflection. Relation between length scale and timescale for observing long time steady $\alpha$ value.  \textbf{Bottom}: How $\alpha_{\rm{longtime}}$ converges towards 1 with increasing system length (\textbf{Right}) and reflection time  (\textbf{Left}).  \textbf{Top Right}: Plot showing length scale and time scale a directly correlated in both w/ eq. and w/o eq. initializations.}
\end{center}
\end{figure}

\begin{figure}[!ht]
\begin{center} 
    \resizebox{.5\textwidth}{!}{\includegraphics{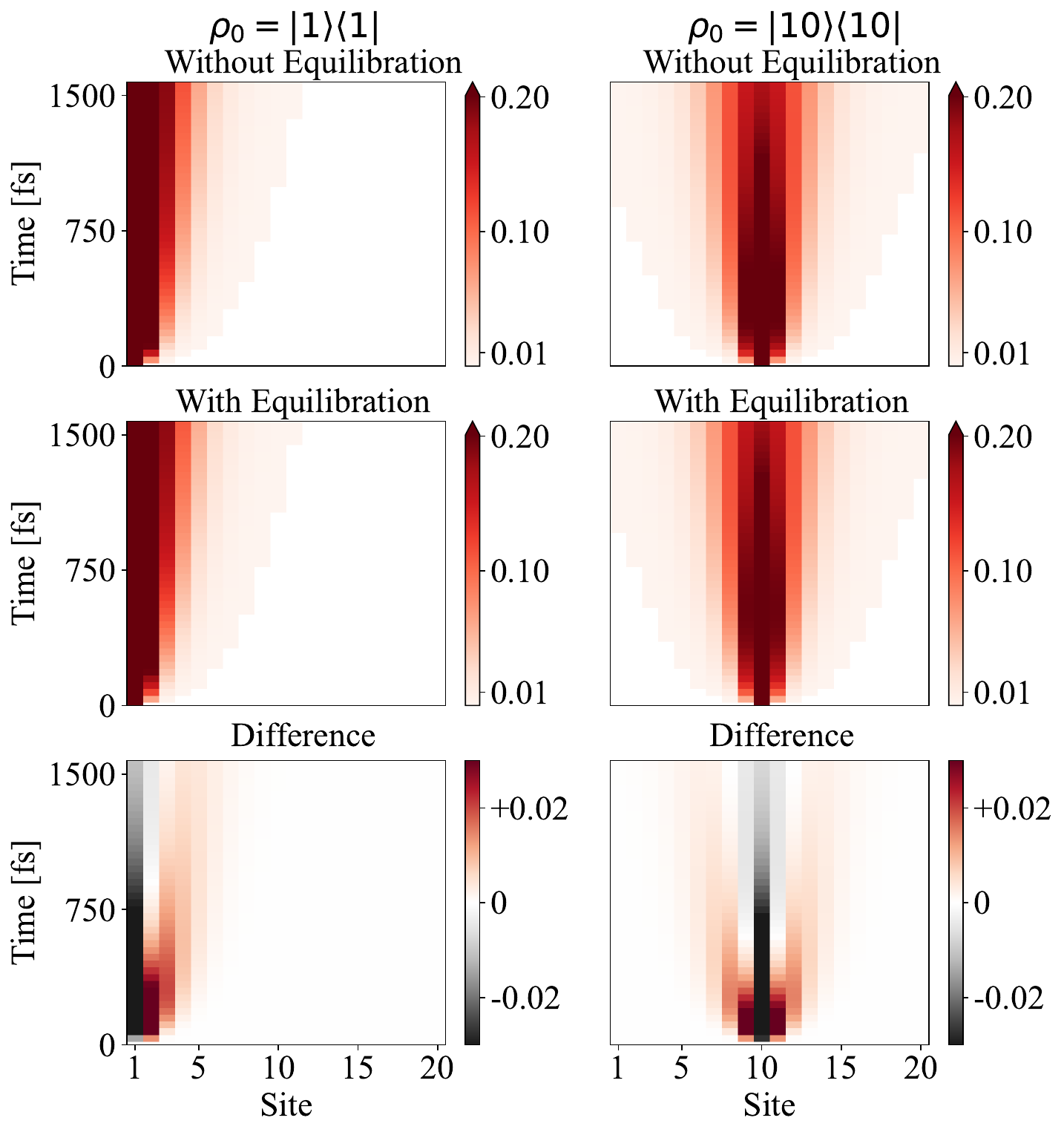}}
\vspace{-20pt}
\caption{\label{fig:heatmap_linear_si} Nonequilibrium HEOM results for a 20-site open chain. These calculations were performed with $dt = 0.25$~fs, $\eta = 323$~cm$^{-1}$, $\gamma = 41$~cm$^{-1}$, and $v_{ij} = 50$~cm$^{-1}$. The dynamic charge density as a function of lattice site for both types of nonequilibrium simulation for both initial conditions. The difference between the top two panels confirms the initial polaron relaxation effect decreases with time.}
\end{center}
\end{figure}

\begin{figure}[!ht]
\begin{center} 
    \resizebox{.5\textwidth}{!}{\includegraphics{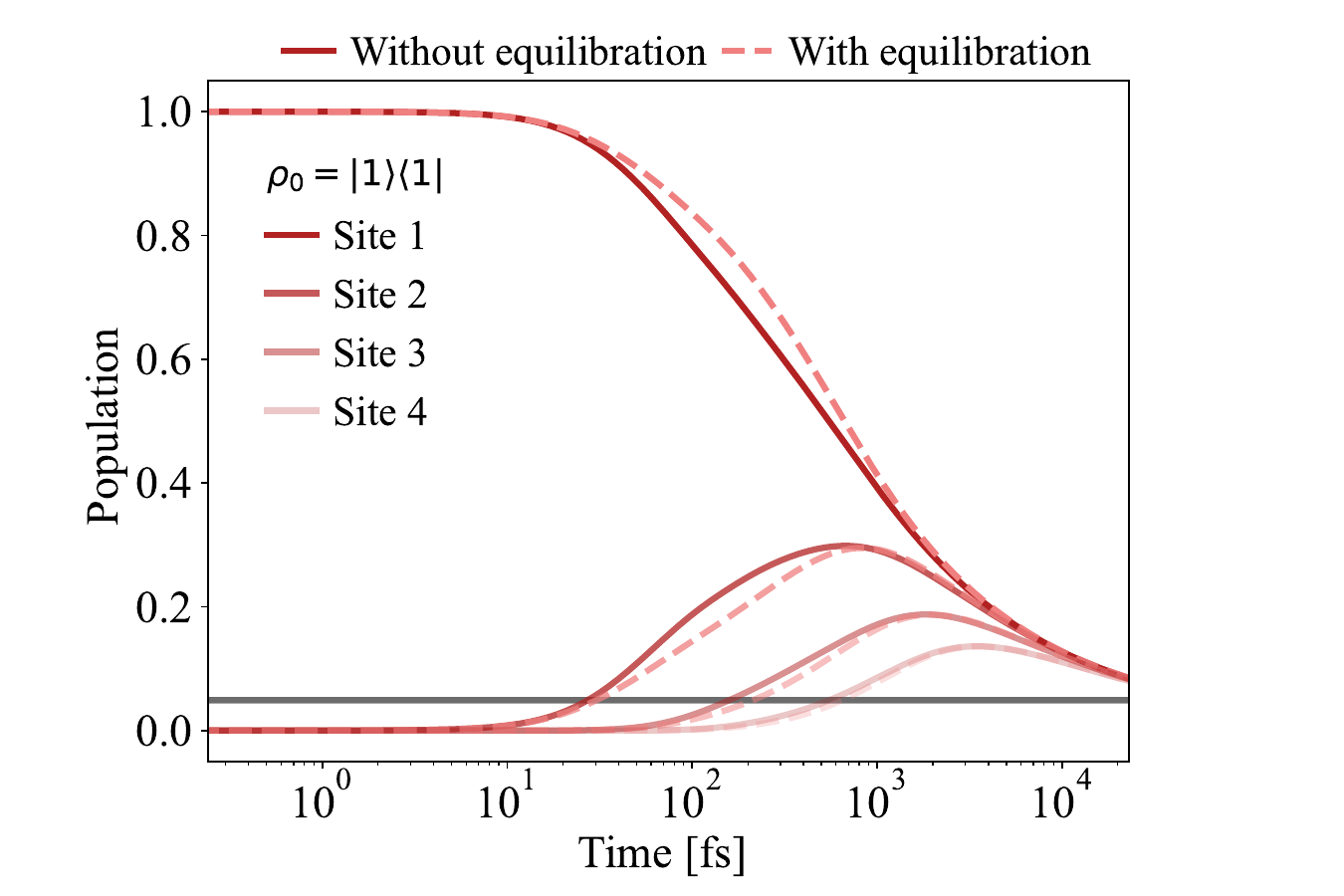}}
\vspace{-20pt}
\caption{\label{fig:pop_persite_time}  Nonequilibrium HEOM results for a 20-site open chain with $dt = 0.25$~fs, $\eta = 323$~cm$^{-1}$, $\gamma = 41$~cm$^{-1}$, and $v_{ij} = 50$~cm$^{-1}$. Initializing the charge next to the wall in an open chain configuration,  the population of each plotted site approaches the equilibrium value (solid grey line) from above.}
\end{center}
\end{figure}

\begin{figure}[!ht]
\begin{center} 
    \resizebox{.5\textwidth}{!}{\includegraphics{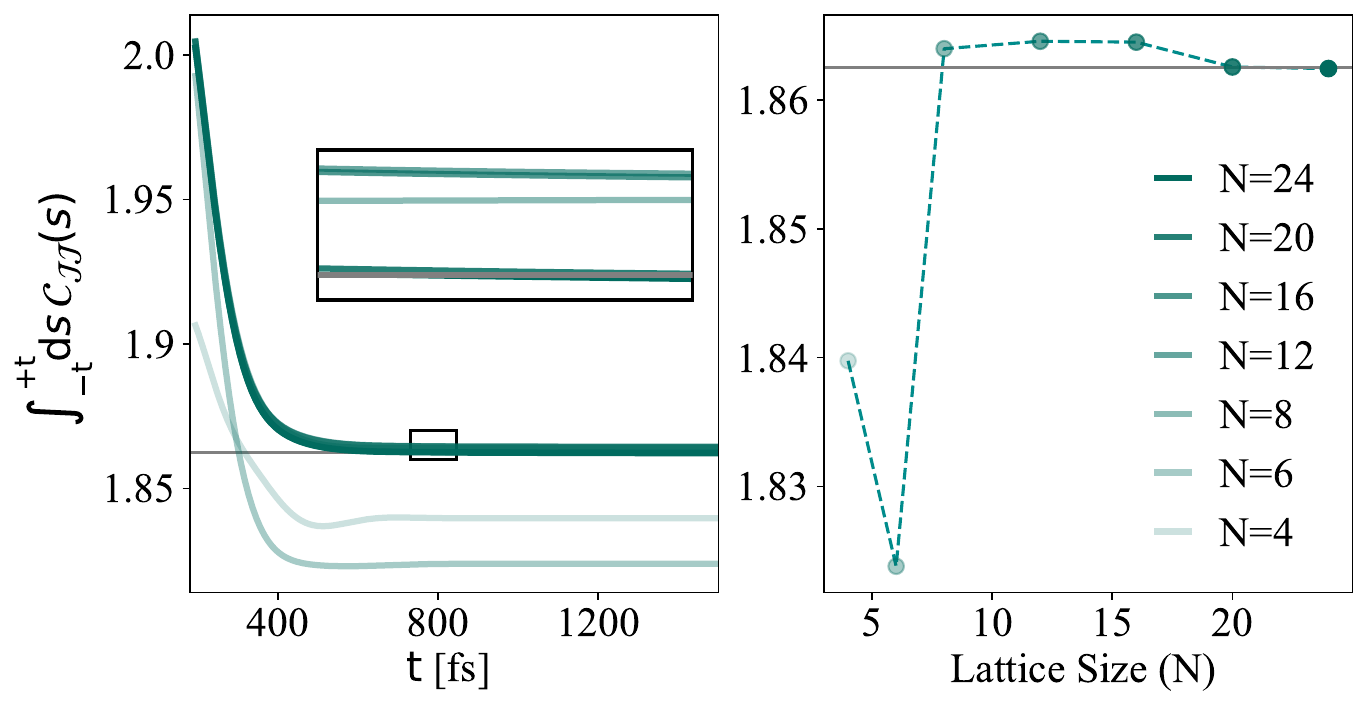}}
\vspace{-16pt}
\caption{\label{fig:conveged_eqm_simulation} Equilibrium simulation converged with time and system size: Equilibrium HEOM results for the periodic topology. The dynamics were calculated with $dt = 0.25$~fs, $v_{ij} = 50$~cm$^{-1}$, $\eta = 30$~cm$^{-1}$, and $\gamma= 25$~cm$^{-1}$. \textbf{Left}: Integral of $C_{JJ}$ against its integration limit for different periodic chain lengths. \textbf{Inset}: Zoom of the region indicated by the black box $\sim 800$~fs, showing the $N>6$ curves in detail. \textbf{Right}: Convergence plot of the integral against ring size (final points of the left panel). The grey dashed line in both panels shows the final, converged value as being 1.862~\AA$^2$fs$^{-1}$.}
\end{center}
\end{figure}

\clearpage

\end{document}